\documentclass[12pt]{article}

\usepackage{epsfig,amsfonts,newlfont,rotate}

\DeclareMathAlphabet{\mathitbf}{T1}{cmr}{bx}{it}

\newcommand{\e}{\mathrm e}
\newcommand{\bbox}[1]{\mbox{\boldmath $#1$}} 

\begin{document}

\title{On the Universality Class of Monopole Percolation\\ in Scalar QED}
\author{L.~A.~Fern\'andez and V.~Mart\'{\i}n-Mayor,\\
\normalsize \it Departamento de F\'{\i}sica Te\'orica I, 
         Facultad de CC. F\'{\i}sicas,\\  
\normalsize  Universidad Complutense de Madrid, 28040 Madrid, Spain.\\
{\small e-mail: \{{\tt laf,victor}\}{\tt @lattice.fis.ucm.es}}
}
\date{December 7, 1998}

\maketitle

\begin{abstract}
\small We study the critical properties of the monopole-percolation
transition in U(1) lattice gauge theory coupled to scalars at infinite
($\beta=0$) gauge coupling.  We find strong scaling corrections in the
critical exponents that must be considered by means of an
infinite-volume extrapolation.  After the extrapolation, our results
are as precise as the obtained for the four dimensional
site-percolation and, contrary to previously stated, fully compatible
with them.
\end{abstract}
  
\vskip 5 mm

Monte Carlo.
Percolation.
Critical exponents.
Phase transitions.
Finite-size scaling.
QED.


\section{Introduction}

The richness of the phase diagram of models containing charged scalars
and/or fermions and abelian gauge fields has produced the hope of
finding a non-trivial critical point in four dimensions.  For the
compact formulation of lattice QED, the different phases of these
models (Confining, Higgs, Coulomb), can be characterized in terms of
their topological content~\cite{TOPOLOGIA}, and this relation can be
numerically explored~\cite{TOPOLOGIANUM} (see Ref.~\cite{FKL} for a
detailed exposition). The link between topology and phase-diagram is
strong to the point that a monopole-percolation second-order phase
transition can be found {\em beyond} the end-point of the
Higgs-Coulomb phase transition line (see Ref.~\cite{U1HIGGS} for a
study of the phase-diagram). In Ref.~\cite{KOGUTCONJ}, it has been
conjectured that the monopole-percolation phase transition can produce
a chiral phase transition that, being driven by the monopole
percolation, would present the same critical exponents.

This conjecture has been put to test in Ref.~\cite{FKL}, where the
position of both the monopole-percolation and chiral critical lines
have been located, by means of a Monte Carlo simulation in the
quenched approximation. The chiral critical line is very close to the
monopole-percolation one, but they can be clearly resolved in the
limit of infinite ($\beta=0$) gauge coupling. A clear-cut test of the
scenario proposed in Ref.~\cite{KOGUTCONJ} would be an accurate
measure of both the chiral and monopole-percolation critical
exponents.  The critical exponents for monopole-percolation have been
measured in Refs.~\cite{FKL,BAIG}. The critical exponents displayed a
very mild variation along the critical line, consistent with a single
Universality Class, although significantly different from the
site-percolation~\cite{PERC4D}.  However, in Refs.~\cite{FKL,BAIG} the
scaling corrections are not considered in the analysis.

In this letter, we report the results of a Monte Carlo calculation of
compact scalar QED in the strong coupling limit for the gauge field
($\beta=0$), where the model is integrable. In this way, we are able
to directly generate independent configurations obtaining accurate
measures in large lattices.

In our study, we shall use a Finite-size Scaling (FSS) method based on
the comparison of measures taken in two lattices, at the coupling
value for which a renormalization group invariant (namely the
correlation length in units of the lattice size) takes the same value
in {\em both} lattices~\cite{PERC4D,OURFSS,ISINGPERC}. In this way, by
considering the scaling of other dimensionless quantities, we obtain
direct information on corrections to scaling. In the present case the
scaling corrections are notoriously difficult to deal with. The usual
strategy of just considering the leading scaling corrections would
only work in extremely large lattices, and we are compelled to use
sub-leading corrections in the infinite volume extrapolation.  After
this extrapolation, we conclude that monopole-percolation belongs to
the same Universality Class that site-percolation.

\section{The model}

The action for the (compact) U(1)-gauge model on the lattice, coupled
to unit modulus scalars can be written as
\begin{equation}
S=-\beta\sum_{\mathitbf{r},\mu<\nu}\mathrm{Re}\, U_{\mu\nu}(\mathitbf{r})
  -\kappa\sum_{\mathitbf{r},\mu}\mathrm{Re}\, \Phi^\dagger(\mathitbf{r})
U_\mu(\mathitbf{r})\Phi(\mathitbf{r}+{\bbox{\mu}})\ ,
\label{ACTION}
\end{equation}
where $\mathitbf r$ is the four-dimensional lattice site, $\mu$ and
$\nu$ run over the four spatial directions, ${\bbox \mu}$ is the
vector joining neighbours along the $\mu$ direction,
$U_{\mu\nu}(\mathitbf{r})$ is the elementary plaquette,
$U_\mu(\mathitbf{r})$ the gauge variable, and $\Phi(\mathitbf{r})$ the
scalar field.  The lattice volume is $V=L^4$ and periodic boundary
conditions are imposed.  Notice that we use the normalization of the
parameter $\kappa$ as in Ref.~\cite{U1HIGGS} which is twice that used
in Ref.~\cite{FKL}.

For $\beta=0$ and after a gauge transformation to eliminate the
$\Phi$ fields, the action simplifies to
\begin{equation}
S=-\kappa\sum_{\mathitbf{r},\mu}\mathrm{Re}\, U_{\mu}(\mathitbf{r})\ .
\label{ACTIONFIN}
\end{equation}
The generation of independent Monte Carlo configurations for this
action is straightforward, as the link variables are dynamically
independent.

To study the monopoles in the lattice, let us write
$U_{\mu}(\mathitbf{r})=\e^{\mathrm{i} \theta_\mu(\mathitbf{r})}$ and define
\begin{equation}
\theta_{\mu\nu}(\mathitbf r)=
\theta_\mu(\mathitbf r)
+\theta_\nu(\mathitbf r+{\bbox\mu})
-\theta_\mu(\mathitbf r+{\bbox\nu})
-\theta_\nu(\mathitbf r)=
\overline \theta_{\mu\nu}(\mathitbf r)+2\pi N_{\mu\nu}(\mathitbf r)\ ,
\end{equation}
where $\overline \theta_{\mu\nu}$ is taken in the interval
$(-\pi,\pi]$ and $N_{\mu\nu}$ is an integer.

We obtain the monopole current in the dual lattice as~\cite{TOPOLOGIANUM}
\begin{equation}
m_\mu(\tilde\mathitbf r)=\frac{1}{2}\epsilon_{\mu\nu\rho\sigma}\Delta^+_\nu
N_{\rho\sigma}(\mathitbf r+{\bbox \mu})\ ,
\end{equation}
where $\Delta^+$ is the forward difference operator in the lattice.
Each component of the current $m_\mu$ is an integer which lives in the
link of the dual lattice leaving $\tilde\mathitbf r$ in the ${\bbox
\mu}$ direction.  Clusters are defined as sets of sites of the dual
lattice connected through links with nonzero monopole current ({\em
occupied} links).

The observables that we measure for every gauge configuration are the
link energy and {\em magnetization-like} quantities that can be
expressed in terms of the cluster-size distribution:
\begin{equation}
\begin{array}{lcl}
E&=&\displaystyle\sum_{\mathitbf{r},\mu}\mathrm{Re}\, 
        U_{\mu}(\mathitbf{r})\ ,\\
M_1&=&\displaystyle\sum_c n_c\ , \\
M_2&=&\displaystyle\sum_c n_c^2\ , \\
M_4&=&\displaystyle3M_2^2-2\sum_c n_c^4\ , \\
\displaystyle M_{\mathrm{max}}&=&\displaystyle\max_c n_c\ ,
\end{array}
\end{equation}
where $n_c$ is the number of sites of the $c-$nth cluster. $E$ is used
to compute $\kappa$ derivatives and to extrapolate the measures taken
at $\kappa$ to neighbouring coupling values~\cite{REWEIGHT}.  The
definition of $M_2$ ($M_4$) can be understood by putting in the
occupied sites Ising spins at zero temperature (spins on the same
cluster have the same sign), taking the second (fourth) power of the
magnetization, and averaging over the signs of the
clusters~\cite{PERC4D}.

The same construction allows for a sensible measure of the
correlation length in a finite lattice. We first measure the Fourier
transform of the clusters
\begin{equation}
\widehat n_c(\mathitbf{k})= 
\sum_{\mathitbf{r}\in c} e^{\mathrm i \mathitbf{k}\cdot\mathitbf{r}}\, 
\end{equation}
at minimal momentum, from which we obtain
\begin{equation}
F=\frac{1}{4}\left\langle \sum_{\Vert\mathitbf{k}\Vert=2\pi/L} 
        \sum_c |n_c(\mathitbf{k})|^2 \right\rangle\ ,
\end{equation}
and then use the following definition~\cite{COOPER}
\begin{equation}
\xi=\left(\frac{\langle M_2\rangle/F-1}{4\sin^2(\pi/L)}\right)^{1/2}\ .
\end{equation}

We have used two definitions of the
unconnected susceptibility
\begin{equation}
\begin{array}{lcl}
\chi_1&=&\langle M_2 \rangle /V\ , \\
\chi_2&=&\langle M_{\mathrm{max}}^2 \rangle /V\ .
\end{array}
\label{CHIS}
\end{equation}
Notice that the monopole density $M_1/V$ is not critical at the
transition. So, we can define also the susceptibilities dividing $M_2$
or $M_\mathrm{max}$ by $M_1$, as done in Refs.~\cite{FKL,BAIG}. This
should only modify the corrections to scaling (in general, we find
that they increase slightly).

The previous numerical studies have been mainly based on measures of
connected susceptibilities, specifically
\begin{equation}
\begin{array}{lcl}
\overline\chi_1^\mathrm{c} &=&\langle (M_2-M_{\mathrm{max}}^2)/M_1 \rangle\\
\overline\chi_2^\mathrm{c}&=&(\langle (M_{\mathrm{max}}/M_1)^2 \rangle- 
\langle M_{\mathrm{max}}/M_1 \rangle^2) V\ .
\end{array}
\end{equation}
These definitions make sense at both sides of the transition and
present a peak near it. However we shall see that they present strong
corrections to the scaling, becoming less appropriate than the
unconnected ones for a FSS study.

It is also very useful to measure quantities that keep bounded at the
critical point, but whose $\kappa$ derivatives diverge. Some examples
are the correlation length in units of the lattice size and the Binder
parameters:
\begin{equation}
\begin{array}{rcl}
B_1&=&\displaystyle\frac{1}{2}\left(3-\frac{\langle M_4\rangle}
        {\langle M_2\rangle^2}\right)\ ,\smallskip\\ 
B_2&=&\displaystyle\frac{\langle M_2\rangle}{\langle
        M_{\mathrm{max}}\rangle^2}\ ,\smallskip\\
B_3&=&\displaystyle\frac{\langle M^2_{\mathrm{max}}\rangle}{\langle
        M_{\mathrm{max}}\rangle^2}\ .
\end{array}
\end{equation}

\section{The numerical method}

We have simulated in symmetric lattices of linear sizes
$L=6,8,12,16,24,32$, and $48$.  We have generated $10^6$ independent
configurations for each lattice. The statistical analysis have been
done with 1000 bins of data, for an accurate error determination.  To
speed up the computations, we have used the U(1) subgroup
$Z_{65535}$. We have checked that the U(1)-discretization effects are
negligible comparing the results with those using $Z_{255}$.

In order to measure critical exponents, we use a FSS
method. Specifically, we use the quotients method used in
Refs.~\cite{PERC4D,OURFSS,ISINGPERC}. Given an observable $O$ that
diverges as $t^{-x_O}$, $t$ being the reduced temperature
$(\kappa-\kappa_{\mathrm c})/\kappa_{\mathrm c}$, the FSS ansatz
predicts that for a finite lattice of size $L$, in the critical region
\begin{equation}
\langle O(L,t) \rangle =L^{x_O/\nu}[F_O(\xi(L,t)/L)+
O(L^{-\omega})]\ ,
\label{FSS}
\end{equation}
where $F_O$ is a smooth scaling function and $\omega$ is the universal
exponent associated to the leading corrections to scaling. 

To eliminate the unknown function $F_O$, one can compute the quotient
$Q_O$ of the mean value of the observable in two different lattices,
at the coupling value where the correlation lengths in units of the
lattice size is the same (there is a {\em crossing}):
\begin{equation}
\left.Q_O\right|_{Q_\xi=s}=\frac{\langle
O(sL,t_\mathrm{cross})\rangle}
 {\langle O(L,t_\mathrm{cross})\rangle}=s^{x_O/\nu}+O(L^{-\omega})\ ,
\label{QUO}
\end{equation}
lattice sizes being $sL$ and $L$ respectively. The $O(L^{-\omega})$
terms include all powers of $L^{-\omega}$ as well as the corresponding
series produced by sub-leading irrelevant
operators~\cite{ISINGPERC,BARBER,BLOTE}. Other corrections are
generated by the non singular part of the free energy (analytical
corrections). For most observables the analytical corrections are
$O(L^{-\gamma/\nu})$. Let us remark that $Q_O$ and $Q_\xi$ are
statistically correlated, which allows for an important (statistical)
error reduction in Eq.~(\ref{QUO}). In fact, a modification of
Eq.~(\ref{QUO}) can be used to study logarithmic corrections to
mean-field behaviour with rather high accuracy~\cite{ISDIL4D}. Let us
finally remark that a similar two-lattices matching method has also
being extremely successful in lattice QCD studies~\cite{LUSCHER}.

It will be also useful to recall the shift of the {\em finite-lattice
critical point} from the true critical point~\cite{BINDER}:
\begin{equation}
t^{\mathrm {cross}}(L,s)
\propto \frac{1-s^{-\omega}}{s^{1/\nu}-1}L^{-\omega-1/\nu}\ ,
\label{BETACFIT}
\end{equation}
where only the leading scaling corrections have been kept. With a
fixed value of $s$, $t^{\mathrm {cross}}(L,s)\propto
L^{-\omega-1/\nu}$, to be compared with the shift of the peaks of the
connected susceptibility, that goes as $t^{\mathrm{peak}} \propto
L^{-1/\nu}$. These peaks can be measured in a single lattice, but we
loose a factor of $L^{-\omega}$. That is why the quotient method
suffers from smaller scaling corrections. Eq.~(\ref{BETACFIT}) also
applies to the crossing of the Binder cumulants $B_i$.

\section{Results}

\begin{table}[b!]
\begin{center}
\begin{tabular*}{\hsize}{@{\extracolsep{\fill}}rllll}\hline\hline
\multicolumn{1}{c}{$L$} 
& \multicolumn{1}{c}{$\nu$}  
& \multicolumn{1}{c}{$\eta$} 
& \multicolumn{1}{c}{$\nu_{\mathrm{s-p}}$} 
& \multicolumn{1}{c}{$\eta_{\mathrm{s-p}}$} \\\hline
6   &0.6801(34)&$ -0.2336(7)$ & \multicolumn{1}{c}{---}& 
\multicolumn{1}{c}{---} \\
8   &0.6829(36)& $-0.1713(6)$ & \footnotesize 0.689(3)   & \footnotesize
$-0.0687(7)$    \\
12  &0.680(5)  & $-0.1256(6)$ & \footnotesize 0.687(3)   & \footnotesize
$-0.0775(7)$    \\
16  &0.681(6)  & $-0.1095(7)$ & \footnotesize 0.688(4)   & \footnotesize
$-0.0825(6)$    \\
24  &0.689(6)  & $-0.0986(6)$ & \footnotesize 0.691(5)   & \footnotesize
$-0.0868(8)$    \\
\hline\hline
\end{tabular*}
\caption{Critical exponents obtained from Eq.(\ref{QUO}) with
lattice pairs ($L$,$2L$). We have used as operators $\partial
\xi/\partial\kappa$ ($x_{\partial_\kappa\xi}=1+1/\nu$), and 
$\chi_1$
($x_{\chi_1}=\gamma/\nu=2-\eta$). The last two rows display the
corresponding results for site-percolation (obtained with exactly the same
operators) reported in~\cite{PERC4D}.}
\label{TEXPOPERC}
\end{center}
\end{table}

In table~\ref{TEXPOPERC} we show our results for the critical
exponents $\eta$ and $\nu$ as obtained from Eq.~(\ref{QUO}).  The used
operators have been $\chi_1$ ($x_{\chi_1}=\gamma/\nu=2-\eta$) and
$\partial \xi/\partial \kappa$ ($x_{\partial_\kappa\xi}=1+1/\nu$). In
all cases, we consider only the ratio $s=2$. These exponents can be
directly compared with the results for the four dimensional
site-percolation~\cite{PERC4D}. The trend for exponent $\nu$ in both
cases is rather similar: the scaling corrections are significantly
smaller than the statistical errors, so that the results seems stable
with growing $L$. To this accuracy, both measures are compatible.  On
the other hand, for the anomalous dimension, $\eta$, both systems
present significant scaling corrections, although stronger in the
monopole-percolation case.  Therefore, the scaling corrections {\em
must} be dealt with before comparison for $\eta$ can be attempted.

\begin{figure}[t]
\begin{center}
\leavevmode
\rotate[l]{\centering\epsfig{file=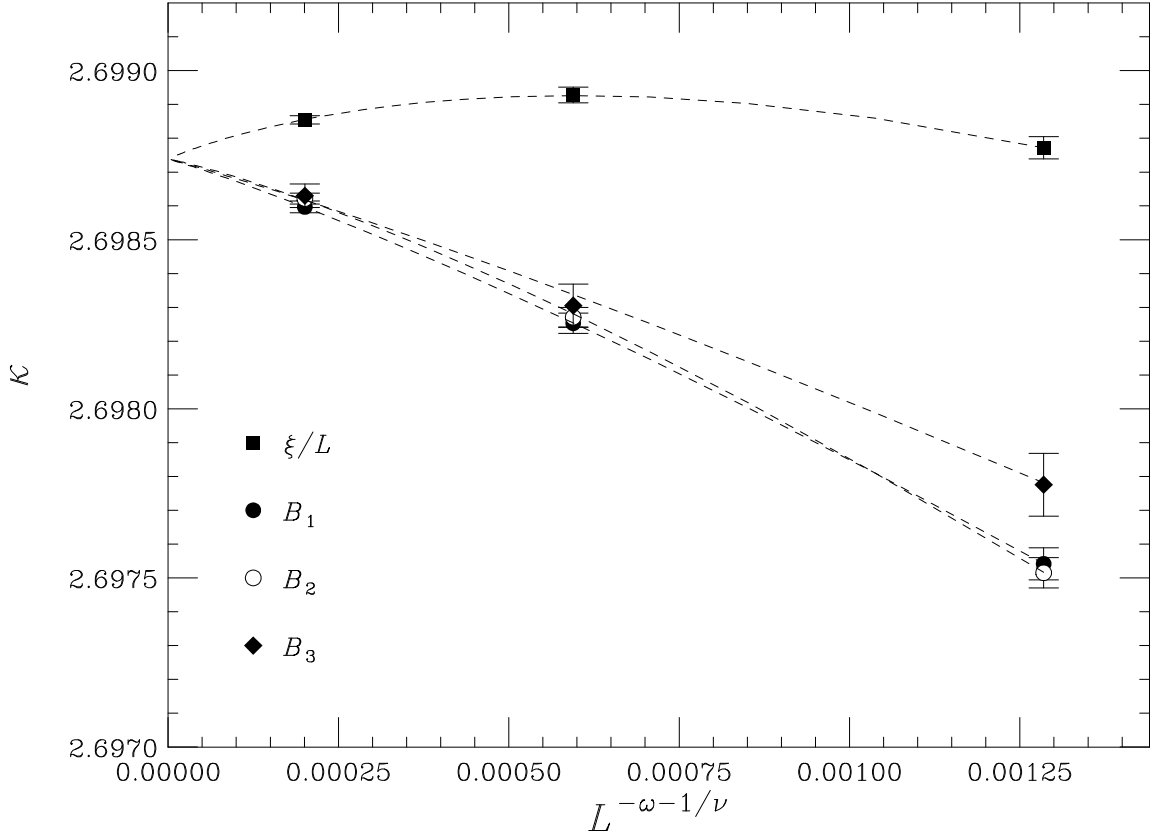,width=0.7\linewidth}}
\end{center}
\caption{Crossing points of
the cumulants, as functions of $L^{-\omega-1/\nu}$ (here we have set
$\omega=1.23$, $\nu=0.69$).}
\label{KAPPAFIG}
\end{figure}

In previous studies~\cite{PERC4D,OURFSS,ISINGPERC}, we have used
Eq.~(\ref{BETACFIT}) to obtain an estimate of $\omega$ that allows to
perform an infinite-volume extrapolation. However, in this case the
higher-order scaling corrections are so large that they need to be
considered. This can be seen in figure~\ref{KAPPAFIG}, where we show
the finite-lattice critical point as obtained by the crossing of
$\xi/L$, and of $B_i$. In the $\xi/L$ case, the behaviour is not even
monotonous with increasing lattice size. Therefore, in this problem we
need to go beyond Eq.~(\ref{BETACFIT}).

Thus, we have considered the three Binder cumulants $B_i$ whose quotients
should behave as
\begin{equation}
\left.Q_{B_i}\right|_{Q_\xi=s}= 1+A_i L^{-\omega}+B_i L^{-2\omega}+
C_i L^{-\omega'} + D_i L^{-\gamma/\nu} + \ldots\ ,
\end{equation}
where $\omega'$ stands for a sub-leading irrelevant exponent.

\begin{figure}[t]
\begin{center}
\leavevmode
\rotate[l]{\centering\epsfig{file=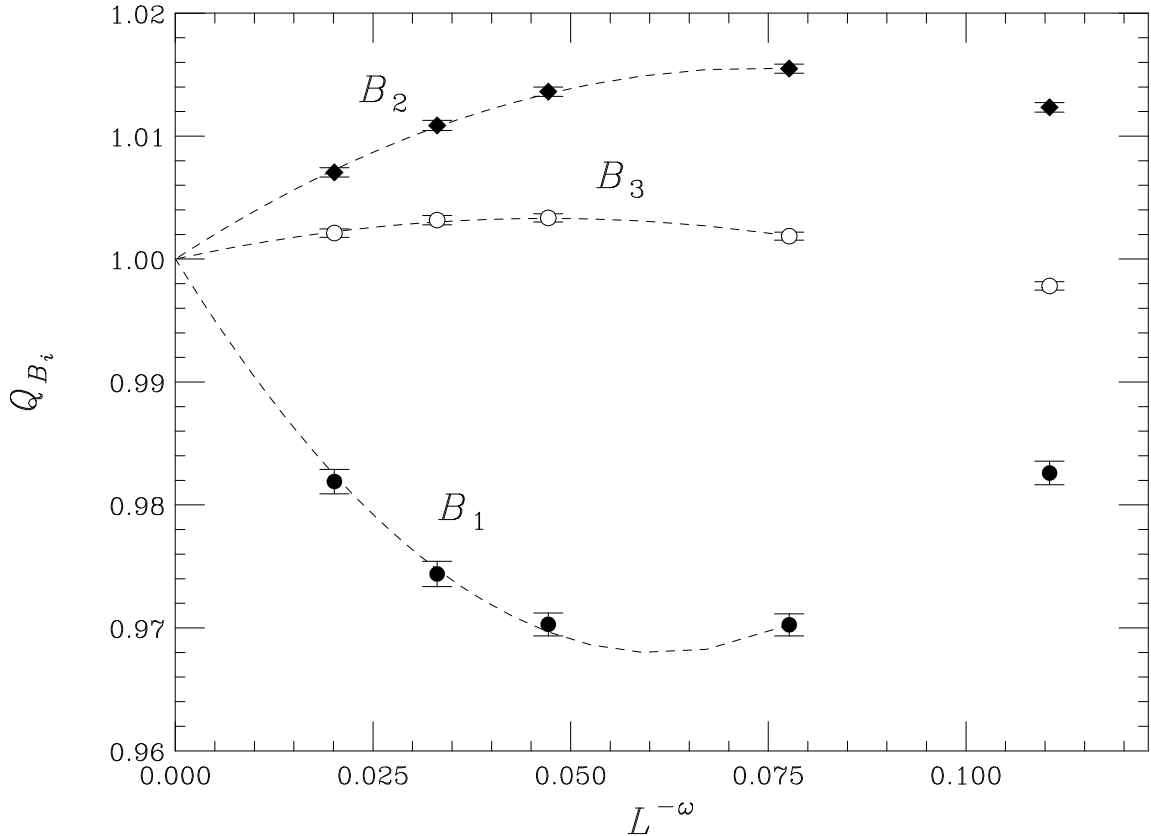,width=0.7\linewidth}}
\end{center}
\caption{Quotients of cumulants $B_1$, $B_2$ and $B_3$ as a function
of $L^{-\omega}$. We use for $\omega$ the value obtained in the fit,
thus the curvature is related with higher-order corrections to scaling.
}
\label{CUMFIG}
\end{figure}

Our numerical results for these quotients are plotted in
figure~\ref{CUMFIG}.  Although the behaviour with growing $L$ is not
monotonous, we have first considered the parametrization $1+A_i
L^{-\omega}$, which should be adequate for large $L$. The quality of
the fit is rather poor unless discarding all but the two largest
pairs: $\omega=0.81(2)\ (\chi^2/\mathrm{dof}=159/8)$; $\omega=0.99(5)\
(\chi^2/\mathrm{dof}=19.9/5)$; $\omega=1.03(8)\
(\chi^2/\mathrm{dof}=5.2/2)$, for $L\geq 8$, $L\geq 12$ and $L\geq 16$
respectively (``dof'' is the number of degrees of freedom in the fit).
We see that higher-order corrections need to be included.  As $\omega$
seems slightly larger than one and, from table~\ref{TEXPOPERC} it is
clear that $\gamma/\nu$ will be close to two, $L^{-2\omega}$ and
$L^{-\gamma/\nu}$ will be of the same order.  Therefore, a quadratic
fit in $L^{-\omega}$ should be a good parametrization provided that
there are not sub-leading irrelevant operators in the intermediate
range. In fact, the quadratic fit shown in figure ~\ref{CUMFIG}
(discarding the smaller lattice, $L=6$) is quite reasonable:
\begin{equation}
\omega=1.23(8),\quad \chi^2/\mathrm{dof}=5.4/5 \ .
\label{OMEGA}
\end{equation}
Notice that the data used in the fits are strongly correlated and the
consideration of the full covariance matrix is mandatory.

A cross-check of the determination Eq.~(\ref{OMEGA}) can be done
fitting to the functional form $1+A_i L^{-\omega}+C_i
L^{-\gamma/\nu}$, fixing $\gamma/\nu=2.09$ (this is the value found in
Ref.~\cite{PERC4D}, for site-percolation and it is also quite close to
the values encountered in table~\ref{TEXPOPERC}). This fit yields a
compatible value with a rather increased error: $\omega=1.36(14)$.

Therefore, our estimate seems consistent although one may prefer
to double the error in Eq.~(\ref{OMEGA}), as an estimation of
systematic errors, to be in the safe side.

A different cross-check is the infinite-volume extrapolation for
$\kappa_{\mathrm c}$, using Eq.~(\ref{BETACFIT}) and the crossing
points for $\xi/L , B_1 ,B_2$ and $B_3$.  In figure \ref{KAPPAFIG} we
plot the cumulant crossing points as a function of
$L^{-\omega-1/\nu}$. As the behaviour is not monotonous we have tried
a fit to $\kappa_{\mathrm{c}}+A_iL^{-\omega-1/\nu}+B_i
L^{-2\omega-1/\nu}$.  From table~\ref{TEXPOPERC} and Eq.~(\ref{OMEGA})
we find that $\nu$ can be determined much more accurately than
$\omega$, so its value can be safely fixed to $\nu=0.69$ in the fit.
Fixing also $\omega$ to (\ref{OMEGA}) we obtain an acceptable fit only
if $L\ge 12$ ($\chi^2/\mathrm{dof}=1.1/3$).  We obtain
\begin{equation}
\kappa_\mathrm{c}=2.698736(34)(11)\ .
\end{equation}
Through out the paper the second error will denote the error induced
by the uncertainty in $\omega$.  Therefore, if one chooses to double
the error in $\omega$, this second error needs to be doubled too. As
the value of $\kappa_{\mathrm c}$ will be by far our most precise
result, it is important to check that the discretization effect of
using $Z_{65535}$ instead of U(1) is negligible. In order to do so, we
have repeated the measure of the crossing point for the pair $(12,24)$
using $Z_{255}$ with the same statistics, obtaining a compatible value
within errors (one per million).

\begin{figure}[t]
\begin{center}
\leavevmode
\rotate[l]{\centering\epsfig{file=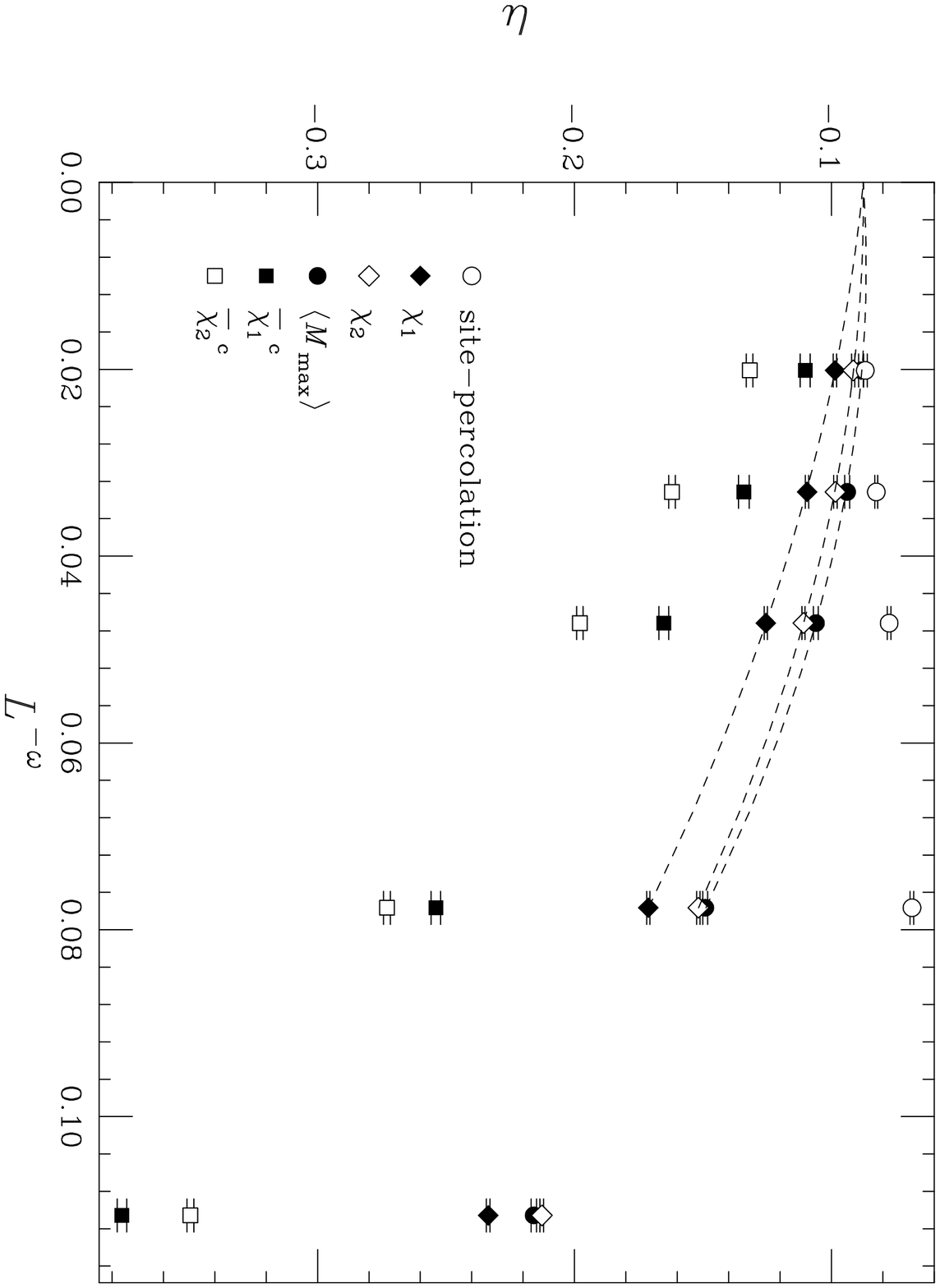,width=0.7\linewidth}}
\end{center}
\caption{The exponent $\eta$ as a function of $L^{-\omega}$ for
several magnetization or susceptibility operators. The dashed lines
are quadratic fits in the variable $L^{-\omega}$ ($\omega=1.23$).
Notice that the connected susceptibilities present stronger scaling
corrections.  We also plot the results of the site-percolation taken
from Ref.~\cite{PERC4D}.}
\label{ETAFIG}
\end{figure}

The exponent $\eta$ can be obtained from the susceptibilities $\chi_1$
and $\chi_2$ (which scale as $L^{\gamma/\nu}=L^{2-\eta}$) and the
magnetization $\langle M_\mathrm{max} \rangle$ (which scales as
$L^{-\beta/\nu+4}=L^{3-\eta/2}$). In figure~\ref{ETAFIG}, we show the
three $\eta$ determinations. It is clear from the plot that we cannot
keep just the leading scaling corrections.  In this figure we show a
joint fit of the three quantities, quadratic in $L^{-\omega}$, with
$L\ge 8$. The extrapolated value is
\begin{equation}
\eta=-0.0876(22)(6),\quad \chi^2/\mathrm{dof}=1.2/5\ .
\end{equation}
This might be compared with the result obtained by assuming that
the scaling corrections are $A_iL^{-\omega} + C_iL^{-\gamma/\nu}$
(fixing again $\gamma/\nu=2.09$):
\begin{equation}
\eta=-0.0902(25)(1),\quad \chi^2/\mathrm{dof}=2.4/5\ .
\end{equation}
We see that by following our recipe of doubling the $\omega$ induced
error, both extrapolations are not covered by the second error bar. We
thus take the difference to estimate the systematic error involved in
the infinite-volume extrapolation generated by higher-order terms. The
final value that includes both kind of errors is
\begin{equation}
\eta=-0.089(4)\ .
\end{equation}

For computing the $\nu$ exponent we measure the quotient corresponding
to the operator $\partial \xi/\partial \kappa$, which scales as
$L^{1+1/\nu}$. From table~\ref{TEXPOPERC} we see that the
infinite-volume extrapolation is not so crucial in this case. However,
we cannot just average the different determinations (which is
basically what a log-log fit does), as the statistical error in the
mean can decrease enough to uncover the scaling corrections. In order
to obtain a safe error estimate, we perform a fit linear in
$L^{-\omega}$. The difference between the extrapolation with $L\ge 6$
and with $L\ge 8$ is ten times smaller than the error. Taking the
error from the latter we obtain:
\begin{equation}
\nu=0.685(6),\quad \chi^2/\mathrm{dof}=1.0/3\ .
\end{equation}
In this case, the $\omega$-induced error is twenty times smaller than
the statistical error.

We summarize the infinite-volume extrapolation of critical exponents
and of the critical coupling in table \ref{COMPARACION}.

\begin{table}[b!]
\begin{center}
\begin{tabular*}{\linewidth}{@{\extracolsep{\fill}}lllll}\hline\hline
\multicolumn{1}{c}{Source} 
& \multicolumn{1}{c}{$\kappa_\mathrm{c}$} 
& \multicolumn{1}{c}{$\nu$} 
& \multicolumn{1}{c}{$\eta$} 
& \multicolumn{1}{c}{$\omega$} \\ \hline
This work &2.69874(6)&0.685(6) &$-0.089(4)$&1.23(16)\\
Franzki, Kogut, Lombardo~\cite{FKL} &2.6938(8) &0.61(4)  &$-0.28(2)$ 
&\multicolumn{1}{c}{---}\\
Site-percolation~\cite{PERC4D}&\multicolumn{1}{c}{---}
        &0.689(10)&$-0.094(3)$&1.13(10)\\
\hline\hline
\end{tabular*}
\caption{Summary of the values obtained for the critical coupling
including statistical and our estimation of systematic errors. In the
second row we show the results of Ref.~\cite{FKL} for the same system.
Finally (third row) we recall the results of Ref.~\cite{PERC4D} for
the site-percolation.}
\label{COMPARACION}
\end{center}
\end{table}

\section{Conclusions}

We have studied with Finite-size Scaling techniques the monopole
percolation transition of compact QED coupled to scalars in the
strong-coupling limit ($\beta=0$). We have shown that state of the art
techniques for measuring critical exponents in spin models can be
successfully applied to this problem.  The approach relies in
comparison of measures taken in two different lattices when a matching
condition is fulfilled (see Eq.~(\ref{QUO})). The efficiency of the
method is greatly enhanced by the availability of a re-weighting
method~\cite{REWEIGHT}, and of an easily measurable Renormalization
Group invariant (the correlation length in units of the lattice
size~\cite{COOPER}).

With the achieved accuracy in individual measures, the scaling
corrections for the reachable lattice sizes are big enough to require
the consideration of sub-leading scaling-corrections in the
infinite-volume extrapolation. After this extrapolation, it is found
(within errors) that at $\beta=0$ monopole-percolation and
site-percolation belong to the same Universality Class. The
discrepancy with previous calculations is explained by the presence of
strong corrections to scaling.

The present study can be extended to the monopole-percolation critical
line at non-zero $\beta$ coupling~\cite{FKL,BAIG}, although the number
of independent configurations that could be generated would be quite
smaller. Another interesting matter is the comparison with the chiral
critical behaviour. Our method would be useful in this respect only if
it is found an analogous of the correlation length in a finite
lattice, that made sense at (chiral) criticality.

\section*{Acknowledgments}
This work has been partially supported by CICyT AEN97-1708 and AEN97-1693.
We gladly acknowledge interesting discussions with J.J. Ruiz-Lorenzo.
\hfill

\end{document}